\documentclass[journal=jpccck, manuscript=article]{achemso}

\usepackage{amsmath, amssymb}
\usepackage{multirow}
\usepackage{array}
\usepackage{textcomp, gensymb}

\author{Francesco Tumino}
\email{francesco.tumino@queensu.ca}
\affiliation{Department of Chemistry, Queen's University, 90 Bader Lane, K7L 2S8 Kingston ON, Canada}
\alsoaffiliation{Department of Energy, Politecnico di Milano, Via Giuseppe Ponzio 34/3, 20133 Milano, Italy}
\author{Sergio Tosoni}
\affiliation{Dipartimento di Scienza dei Materiali, Università di Milano-Bicocca, Via Roberto Cozzi 55, 21025 Milano, Italy}
\author{Paolo D'Agosta}
\author{Valeria Russo}
\author{Carlo E. Bottani}
\author{Andrea Li Bassi}
\author{Carlo S. Casari}
\affiliation{Department of Energy, Politecnico di Milano, Via Giuseppe Ponzio 34/3, 20133 Milano, Italy}

\title[]
{Surface Sensitive Raman Response of Metal-Supported Monolayer MoS$_2$}


\begin{document}
	
	\begin{tocentry}
		\includegraphics[]{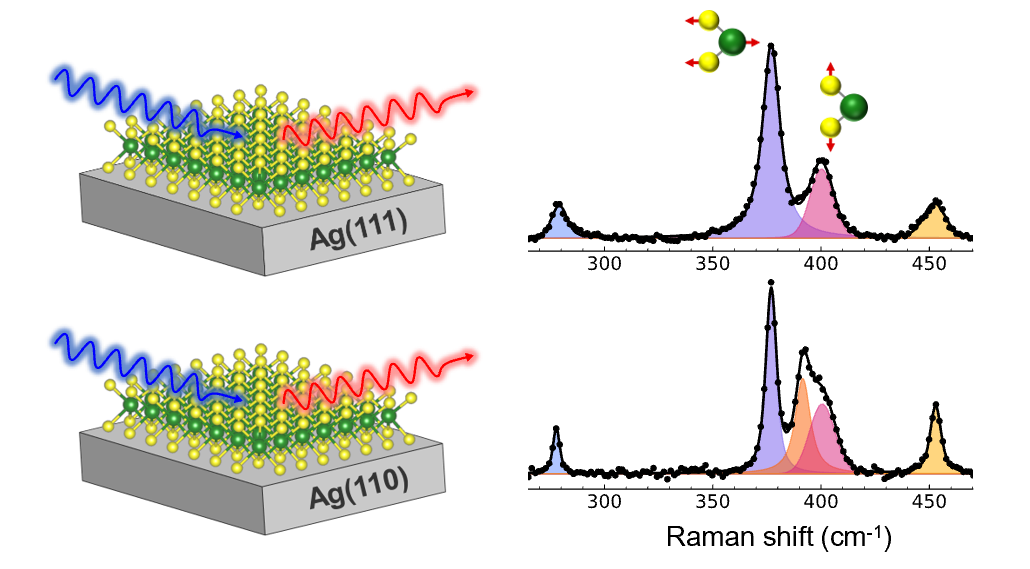}
	\end{tocentry}
	
	\begin{abstract}
		The Raman spectrum of monolayer (ML) MoS$_2$ is remarkably affected by the interaction with metals. 
		In this work we studied ML-MoS$_2$ supported by the Ag(111) and Ag(110) surfaces by using a combined experimental and theoretical approach.
		The MoS$_2$ layer was directly grown on atomically clean Ag(111) and Ag(110) surfaces by pulsed laser deposition, followed by in-situ thermal annealing under ultra-high vacuum conditions. 
		The morphology and structure of the two systems were characterized in-situ by scanning tunneling microscopy, providing atomic-scale information on the relation between the MoS$_2$ lattice and the underlying surface.
		Raman spectroscopy revealed differences between the two MoS$_2$-metal interfaces, especially concerning the behavior of the out-of-plane $A'_1$ vibrational mode, which splits into two contributions on Ag(110). 
		The metal-induced effects on MoS$_2$ vibrational modes are further evidenced by transferring MoS$_2$ onto a more inert substrate (SiO$_2$/Si), where the MoS$_2$ Raman response displays a more ``freestanding-like'' behavior.
		The experimental data were interpreted with the support of ab-initio calculations of the vibrational modes, which provided insight into the effect of interface properties, such as strain and out-of-plane distortion.
		Our results highlight the influence of the interaction with metals on MoS$_2$ vibrational properties, and show the high sensitivity of MoS$_2$ Raman modes to the surface structure of the supporting metal.  
	\end{abstract}
	
	
	\section{Introduction}
	Molybdenum disulfide (MoS$_2$) is one of the most promising 2D semiconductors for future applications in electronics and optoelectronics~\cite{wang2012electronics, mak2010atomically, mak2016photonics, radisavljevic2011single}. 
	The development of MoS$_2$-based devices presents several technological challenges including the synthesis of large-area monolayer (ML) MoS$_2$ and the design of functional MoS$_2$-metal contacts. The latter has been the subject of theoretical~\cite{gong2014unusual, kang2014computational, popov2012designing} and experimental investigations~\cite{gong2013metal, sun2014probing} aimed at studying the junction between MoS$_2$ and various transition metals. 
	Interest in MoS$_2$-metal interfaces is continually growing, fueled by the evidence that MoS$_2$ electronic and vibrational properties are significantly influenced by its interaction with metals~\cite{bruix2016single}. 
	This interaction can manifest through strain or charge transfer effects, which are known to affect the electronic and phononic bandstructures~\cite{zhang2015phonon}, as well as the electron-phonon coupling~\cite{chakraborty2012symmetry}.  
	The Raman response of mono- and few-layer MoS$_2$ has been shown to be very sensitive to such effects~\cite{chakraborty2012symmetry, conley2013bandgap, lloyd2016band, rice2013raman}. 
	
	Raman spectroscopy is widely used to estimate the number of layers in few-layer MoS$_2$ films, as the frequency difference between the two most intense Raman modes (namely $E^1_{2g}$ and $A_{1g}$) increases for increasing thickness~\cite{lee2015anomalous, zhang2015phonon, li2012bulk}. 
	However, this relation applies to weakly interacting MoS$_2$ films, such as MoS$_2$ supported by SiO$_2$ or other dielectric substrates, whose Raman response is comparable to that of freestanding MoS$_2$ (we will refer to weakly interacting MoS$_2$ as ``freestanding-like MoS$_2$'').
	In contrast, metal-supported MoS$_2$ shows significant deviations from the freestanding-like behavior. 
	For instance, different reports on Au-supported ML-MoS$_2$ have shown that the in-plane $E'$ and out-of-plane $A'_1$ vibrational modes downshift compared to their freestanding-like frequencies, and that the $A'_1$ mode splits in two contributions of much lower intensity compared to $E'$~\cite{velicky2020strain, pollmann2021large, rodriguez2022activation, yasuda2017out, tumino2019pulsed} (for monolayer MoS$_2$ the notation changes from $E^1_{2g}$ and $A_{1g}$ to $E'$ and $A'_1$, according to the irreducible representation of the $D_{3h}$ symmetry group).
	Moreover, the peak intensity of $A'_1$, which in freestanding-like MoS$_2$ is normally larger than that of $E'$\cite{carvalho2015symmetry, carvalho2016erratum}, significantly decreases relative to the $E'$ mode, causing the inversion of $A'_1$/$E'$ intensity ratio. 
	These anomalies have been attributed to different causes, such as in-plane epitaxial strain~\cite{velicky2020strain} and moiré-modulated out-of-plane strain~\cite{yasuda2017out}, but a comprehensive understanding is still lacking.
	The interface properties play a key role in the anomalous MoS$_2$ Raman response, which has been observed to be influenced by the homogeneity of the contact area~\cite{gong2013metal}, the metal surface roughness, and the presence of interface contaminants~\cite{velicky2018mechanism, tumino2021hydrophilic}.
	Understanding how the interface with metals affects MoS$_2$ vibrational properties is not only important from a fundamental perspective but can also have a notable impact on the diagnostics of metal-MoS$_2$ contacts.   
	
	To gain further insight into this matter, it is important to investigate the relation between the atomic-scale structure of the MoS$_2$-metal interface and the Raman response of ML-MoS$_2$.
	The epitaxial relation between the MoS$_2$ layer and the metal surface determines several interface properties which can affect the MoS$_2$ lattice vibrations, such as in-plane strain, out-of-plane distortion, and charge transfer.
	Therefore, it is reasonable to hypothesize that different facets of the same metal, having different surface structure, may influence the MoS$_2$ vibrational properties in different ways, potentially leaving surface-specific fingerprints on its Raman spectrum.  
	
	To explore this hypothesis, we conducted a combined experimental and theoretical investigation on ML-MoS$_2$ grown on two different low-index Ag surfaces, namely Ag(111) and Ag(110).
	In both cases, MoS$_2$ was grown via pulsed laser deposition (PLD) on atomically flat surfaces under ultra-high vacuum (UHV) conditions. In this way we avoided the effects of surface roughness and contamination, and achieved a clean and homogeneous metal-MoS$_2$ interface.
	The nanoscale structure was characterized via STM, allowing us to obtain atomic resolution images that supported the theoretical modeling of MoS$_2$/Ag interfaces.
	Raman spectroscopy shows that the two samples display different Raman features, with the emergence of a split $A'_1$ mode in MoS$_2$/Ag(110).
	Our theoretical analysis provides insight into the effects of strain, distortion and other interface properties, and corroborates the high surface sensitivity of MoS$_2$ Raman response observed experimentally.
	To further evidence the effects of the interaction with metals, we subsequently transferred MoS$_2$ from Ag onto a SiO$_2$/Si substrate, and observed that its Raman spectrum acquired a more freestanding-like character.
	By discussing the high sensitivity of MoS$_2$ Raman response to the metal surface topology, this work provides deeper insight into the vibrational properties of metal-supported MoS$_2$, opening to further developments in the characterization of MoS$_2$-metal contacts for electronic devices.

	\section{Methods}
	
	\subsection{Experimental Methods}
	Sample preparation and STM were performed under UHV conditions (base pressure in the $10^{-11}$\,mbar range). 
	A monocrystalline Ag(111) film on mica (MaTeck GmbH, Germany) and an Ag(110) single crystal (MaTeck GmbH, Germany) were prepared by several cycles of Ar$^+$ sputtering and annealing to 700\,K, until the surfaces looked clean under STM inspection.
	MoS$_2$ was grown in a dedicated PLD chamber (base pressure in the $10^{-9}$\,mbar range) by ablating a rotating MoS$_2$ target (Testbourne Ltd, UK) with a KrF excimer laser (248 nm wavelength, $\sim$20\,ns pulse duration).
	The target was ablated by focused laser pulses emitted at a rate of 1\,pulse/s. With a nominal pulse energy of 200\,mJ, the fluence on the target averaged to 2\,J/cm$^2$ per pulse. 
	During the ablation the Ag surface was placed 3\,cm away from the target and kept at room temperature (RT). 
	After deposition the sample was annealed to 720\,K for 30\,min in UHV to promote the crystallization of deposited species. 
	The number of laser pulses on the target was tuned between 5 and 10 to produce single-layer MoS$_2$ structures covering approximately 60\% of the surface. This coverage---estimated by analyzing STM images---was found to be sufficient to produce a good signal-to-noise ratio in Raman spectra.
	STM was performed at RT using a VT Omicron microscope in constant-current mode and electrochemically etched W tips. 
	STM images were plane-corrected and analyzed using Gwyddion~\cite{gwyddion}.
	
	Raman spectroscopy was performed ex-situ immediately after taking the sample out of the UHV chamber to minimize possible surface and interface contamination.
	The spectra were collected in backscattering geometry using an InVia Renishaw spectrometer equipped with an Ar laser. 
	We used a 457\,nm (2.71\,eV) excitation, a 2400\,lines/mm diffraction grating, and a $50\times$ objective lens. 
	The spectrometer was calibrated against the 521\,cm$^{-1}$ peak of a Si crystal. 
	The collected Raman spectra were fitted using Voigt functions. 
	
	MoS$_2$ on Ag(111)/mica was then transferred onto 300\,nm thick SiO$_2$/Si via the following wet transfer procedure. 
	First, the sample was left floating on a HCl aqueous solution for 10\,min to facilitate cleaving off the mica substrate. 
	After rinsing the sample in distilled water, the MoS$_2$/Ag film was then peeled off from mica and placed upside down onto SiO$_2$/Si, obtaining an Ag/MoS$_2$/SiO$_2$/Si structure. 
	The sample was heated in air at 373\,K for 5\,min to promote the adhesion of MoS$_2$ to SiO$_2$. 
	A drop of potassium iodide was then applied to the Ag surface to etch away the Ag layer.
	After a few minutes the sample was rinsed in distilled water and let dry in air. 
	The presence of transferred MoS$_2$ flakes was checked by optical microscopy and Raman spectroscopy.     
	
	\subsection{Theoretical Methods}
	Non-spin polarized DFT calculations were carried on with the VASP code~\cite{Kresse1996,Kresse1996b}. Core electrons were described with the Projector-Augmented Wave scheme~\cite{Blochl1994,Kresse1999}, while Mo(4d, 4p, 5s), S(3s, 3p) and Ag(4d, 5s) were treated explicitly. We adopted the Perdew, Burke and Ernzerhof (PBE)~\cite{Perdue1996} functional, corrected for the long-range dispersion with the Becke-Johnson damped D3 scheme~\cite{Grimme2010,damp_BJ_2011}. The MoS$_2$ lattice was relaxed with a kinetic energy cutoff of 600\,eV and a mesh of $6\times6\times6$ K points, yielding a lattice parameter of 3.15\,\AA{} for the free-standing monolayer, in agreement with experimental data~\cite{wildervanck1964preparation, dickinson1923crystal, swanson1953standard}. The relaxation of the Ag lattice, conducted with a kinetic energy cutoff of 600\,eV and a mesh of $12\times12\times12$ K-points, yielded a lattice parameter for the conventional cubic cell of 4.07\,\AA, which agrees with experimental measurements~\cite{davey1925precision, jette1935precision}. 
	For the ionic relaxations of the MoS$_2$/Ag(111) and MoS$_2$/Ag(110) interfaces, the MoS$_2$ lattice was stretched to match that of the underlying Ag, and the ionic positions of the MoS$_2$ monolayer were then fully relaxed, along with those of the three uppermost layers of the Ag slab. The plane waves basis set was expanded within a kinetic energy cutoff of 400\,eV. Convergence thresholds of $10^{-5}$\,eV (electronic loop) and $10^{-2}$\,eV/\AA{} (ionic loop in structural relaxation) were adopted. Dipole and quadrupole corrections to the total energy were applied along the nonperiodic direction. In order to avoid spurious interactions between replica of the slab models, a vacuum region of at least 15\,\AA{} was included in the supercells. Vibrational frequencies were next calculated on the relaxed structure within the harmonic approximation, using the finite difference approach. Only the MoS$_2$ atoms were displaced during the frequency calculations. The phonon dispersion was not considered. The convergence criteria for the self-consistent field loop were tightened to $10^{-7}$\,eV. The off-resonance Raman activity of each normal mode was evaluated by looking at the contribution of the macroscopic dielectric constants to the Raman tensors~\cite{vasp_raman_py}. The out-of-plane component of the dielectric tensor was recalculated by implementing the equations originally proposed by Laturia et al.~\cite{Laturia_dielectric} to account for the 2D nature of the MoS$_2$ monolayer and cancel out the spurious contribution to the dielectric properties of the material arising from the empty region of the supercell~\cite{Laturia_dielectric}. The simulated spectra were generated by summing up a Lorentzian function whose integral matches the Raman intensity of each normal mode, with a half-peak width of 3\,cm$^{-1}$.   
	
	
	\section{Results and Discussion}
	
	\subsection{Morphology and Structure of MoS$_2$ on Ag Surfaces}
\begin{figure}[th]
		\centering
		\includegraphics[width=.85\textwidth]{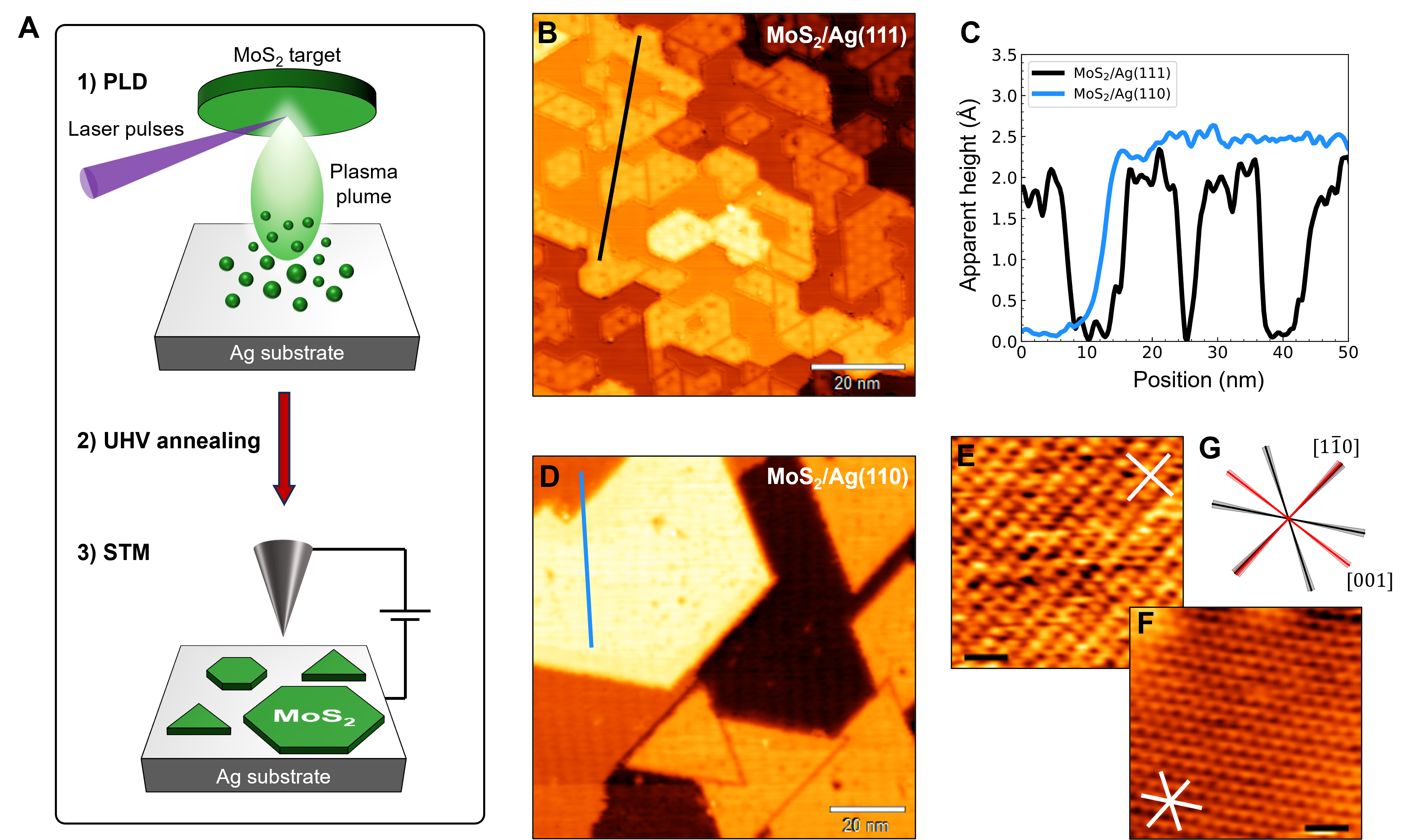}
		\caption{Morphology and structure of PLD-grown MoS$_2$ islands on Ag(111) and Ag(110). (A) Schematic representation of MoS$_2$ fabrication: 1) PLD is used to deposit MoS$_2$ onto freshly prepared Ag surfaces. 2) The sample is then annealed in UHV to favor MoS$_2$ crystallization and 3) observed in-situ via STM at room temperature. (B) Large-scale STM image of MoS$_2$ islands on Ag(111) (1.0\,V, 0.5\,nA). (C) STM height profiles extracted along the lines in B (black profile) and D (blue profile). The measured apparent height of MoS$_2$ islands is compatible with single-layer thick MoS$_2$. (D) Large-scale STM image of MoS$_2$ islands on Ag(110) (1.5\,V, 0.2\,nA). (E-F) Atomic resolution STM images of (E) bare Ag(110) (0.2\,V, 0.2\,nA) and (F) MoS$_2$/Ag(110) (0.06\,V, 0.5\,nA). White lines indicate the close-packing directions of the two lattices. Scale bars: 1\,nm. (G) Comparison between the orientations of MoS$_2$ (black) and Ag(110) (red) lattices. Shaded areas represent the angle uncertainty along lattice directions. MoS$_2$ is aligned along the $[1\bar10]$ direction of the substrate.}
		\label{fig:stm}
	\end{figure}

	Monolayer MoS$_2$ (ML-MoS$_2$) was grown on Ag(111) and Ag(110) using a PLD-based approach that we described in previous studies~\cite{tumino2019pulsed, tumino2021hydrophilic, d2023interface}. This method is capable to produce ML-MoS$_2$ structures on metal surfaces ranging from nanosized islands to polycrystalline 2D layers depending on the number of laser pulses used to ablate the MoS$_2$ target. The UHV compatibility of this approach allows us to characterize in-situ the surface via STM (Figure~\ref{fig:stm}A). 
	Figure~\ref{fig:stm}B shows MoS$_2$ islands on Ag(111), covering $\sim$63\% of the surface. The apparent height of these structures is about 2\,\AA{} (Figure~\ref{fig:stm}C, black line), compatible with the height of metal-supported ML-MoS$_2$ measured by STM in previous works~\cite{sorensen2014structure, tumino2019pulsed, tumino2021hydrophilic}, which typically ranges between 2 and 2.5\,\AA.
	The surface of MoS$_2$ islands shows the characteristic hexagonal moiré pattern due to the lattice mismatch between Ag(111) and MoS$_2$. 
	Previous studies conducted on Au(111) have shown a a similar moiré pattern~\cite{sorensen2014structure, tumino2019pulsed, gronborg2015synthesis, bana2018epitaxial}, which has been described by a 10-MoS$_2$/11-Au coincidence with no rotational mismatch~\cite{bana2018epitaxial}.
	MoS$_2$ on Ag(111) likely produces the same superstructure, due to the structure similarity between the two metal surfaces. 
	This hypothesis is supported by STM measurements recently reported by us~\cite{tumino2021hydrophilic} and other groups~\cite{yousofnejad2020monolayers}, showing a moiré periodicity of 32-33\,nm and a MoS$_2$ lattice constant of 3.15-3.17\,\AA, in agreement with a 10-MoS$_2$/11-Ag coincidence superstructure. 

	Figure~\ref{fig:stm}D shows MoS$_2$ islands on Ag(110). Their apparent height is $\sim$2.5\,\AA{} (Figure~\ref{fig:stm}C, blue line), consistent with single-layer thickness. The islands cover approximately 58\% of the surface, and they appear to be larger than the ones observed on Ag(111). As shown in Figure~\ref{fig:stm}D, MoS$_2$/Ag(110) crystals can grow as large as $\sim$70\,nm in linear size, while on Ag(111) they only reach $\sim$20 nm.        
	The atomic resolution images in Figure~\ref{fig:stm}E-F were taken on a portion of bare Ag(110) and over a MoS$_2$/Ag(110) island, respectively. 
	The lattice constants of MoS$_2$ measured from Figure~\ref{fig:stm}F is $3.2\pm0.2$\,\AA. 
	Figure~\ref{fig:stm}G shows the main symmetry directions of MoS$_2$ (black) and Ag(110) (blue) lattices. The two lattices are aligned along the $[1\bar10]$ direction of the substrate.
	This observation agrees with a recent work by Bignardi et al.~\cite{bignardi2021anisotropic}, where they propose that MoS$_2$ on Ag(110) forms a commensurate superlattice with a periodicity corresponding to five and two times the Ag(110) constants along the $[1\bar10]$ and $[001]$ directions, respectively. 
	STM images like the one reported in Figure S1B show a periodic modulation of MoS$_2$ surface which can be attributed to the moiré pattern formed by the superlattice, in agreement with what reported by Bignardi et al. 
	The commensurate structure of MoS$_2$ on Ag(110) can be described by the $[10/9,\;0\;|\;5/9,\;-2/3]$ overlayer matrix (see SI, Section 1).
	
	\subsection{Raman Spectroscopy of Ag-supported MoS$_2$}
\begin{figure}[ht]
		\centering
		\includegraphics[width=.9\textwidth]{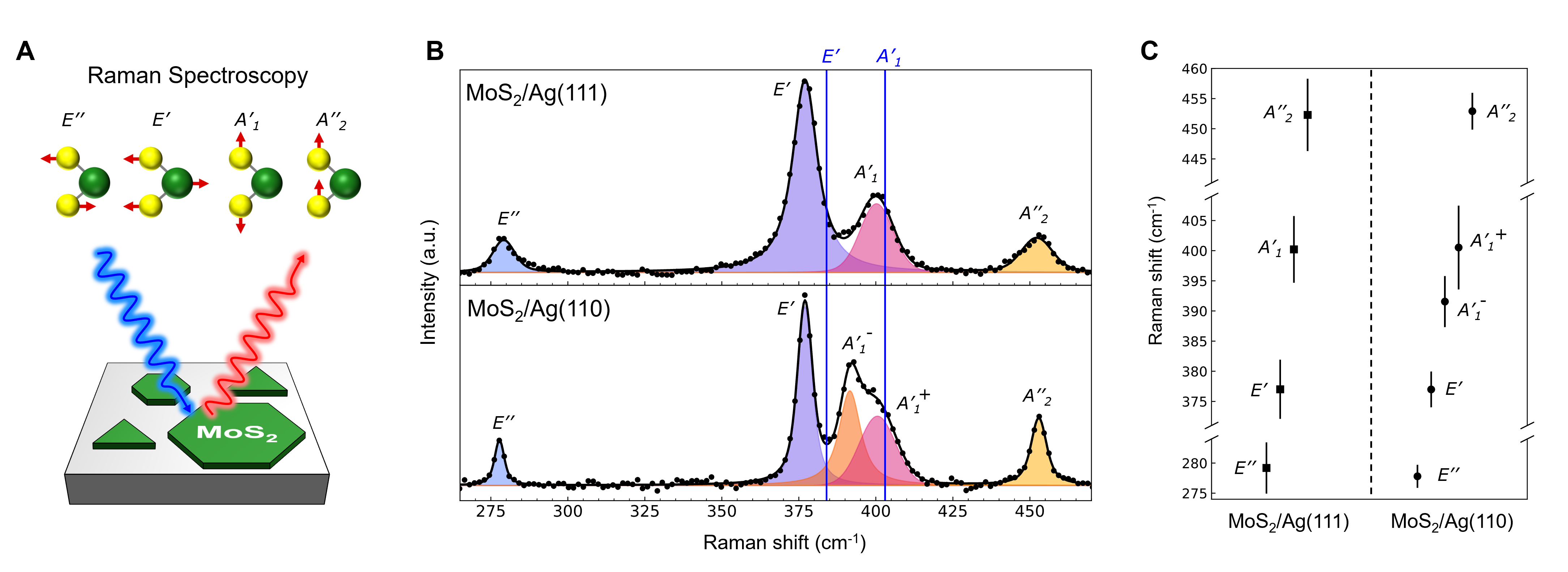}
		\caption{Raman spectroscopy of Ag-supported monolayer MoS$_2$. (A) Schematics illustrating the atomic motions associated to the observed vibrational modes. (B) Raman spectra of MoS$_2$/Ag(111) (top) and MoS$_2$/Ag(110) (bottom), collected with a 457\,nm wavelength. Black dots represent experimental points. The fitting curves (black solid lines) have been obtained by adding together the Voigt functions (colored areas) that fit individual peaks. The blue vertical lines mark the typical positions of $E'$ (384\,cm$^{-1}$) and $A'_1$ (403\,cm$^{-1}$) in ``freestanding-like'' ML-MoS$_2$. (C) Peak frequency (dots) and FWHM (solid lines) of the Raman modes of MoS$_2$/Ag(111) (left column) and MoS$_2$/Ag(111) (right column). Data have been extracted from fitting the experimental spectra in B.}
		\label{fig:raman}
	\end{figure}

	We performed Raman spectroscopy to study the vibrational modes of MoS$_2$ grown on the two Ag surfaces (Figure~\ref{fig:raman}A).  
	Raman spectra were acquired ex-situ as soon as the samples were taken out the UHV chamber. 
	Minimizing the air exposure is important, as we have previously reported~\cite{tumino2021hydrophilic} that contaminants can degrade the purity of the MoS$_2$/Ag and alter the Raman spectrum over a timescale of a few hours.
	
	Figure~\ref{fig:raman}B shows the Raman spectra of MoS$_2$/Ag(111) (top) and MoS$_2$/Ag(110) (bottom) collected using a 457 nm excitation. 
	The spectra have been fitted in the 250--500\,cm$^{-1}$ range using Voigt functions (represented by colored areas in Figure~\ref{fig:raman}B). 
	The blue vertical lines mark the positions of $E'$ and $A'_1$ modes in ``freestanding-like'' ML-MoS$_2$, i.e. when the MoS$_2$ layer is weakly interacting with the substrate, as for example in the case of exfoliated ML-MoS$_2$ supported by SiO$_2$/Si substrates. These frequencies are tipically 384\,cm$^{-1}$ for $E'$ and 403\,cm$^{-1}$ for $A'_1$, as reported in previous works~\cite{conley2013bandgap, li2012bulk, zhang2015phonon}. 
	Table~\ref{tab:exp} reports the peak frequency and FWHM for each of the detected modes in MoS$_2$/Ag spectra, as well as the typical frequencies for freestanding-like MoS$_2$.

	MoS$_2$/Ag(111) (Figure~\ref{fig:raman}B, top panel) shows four peaks which can be attributed to $E''$, $E'$, $A'_1$ and $A''_2$ in order of increasing frequency. 
	The $E''$ and $A''_2$ modes are usually not detected in ML-MoS$_2$, because $E''$ is geometry-forbidden in backscattering and $A''_2$ is not Raman active~\cite{zhang2015phonon}.
	However, as argued in a recent work on the MoS$_2$/Au system~\cite{rodriguez2022activation}, a strong interaction with the substrate can lower the ML-MoS$_2$ symmetry from $D_{3h}$ to $C_{3v}$, resulting in the activation of $E''$ and $A''_2$ (in the $C_{3v}$ point group these modes, as well as the $E'$ and $A'_1$ modes, should be labeled as $E$ and $A_1$, but we will not change the nomenclature to avoid any possible confusion).
	\begin{table}[ht]
		\centering
		\begin{tabular}{| l | c | c | c | c |} 
			\hline
			\textbf{Sample} & $E''$ & $E'$ & $A'_1$ & $A''_2$ \\
			\hline 
			MoS$_2$/Ag(111) & 279.2 (8.5) & 377 (9.8) & 400.2 (11) & 452.3 (12)\\ 
			\hline
			MoS$_2$/Ag(110) & 277.8 (3.8) & 377 (5.9) & 391.6 (8.5), 400.5 (13.9) & 452.9 (6.1)\\
			\hline
			freestanding-like MoS$_2$ & -- & 384 & 403 & --\\
			\hline
		\end{tabular}
		\caption{Peak frequency (FWHM) of the Raman modes detected in the experimental spectra of MoS$_2$/Ag(111) and MoS$_2$/Ag(110) (Figure~\ref{fig:raman}). The third row reports the typical Raman frequencies of the $E'$ and $A'_1$ modes in freestanding-like ML-MoS$_2$ reported in the literature~\cite{conley2013bandgap, li2012bulk, zhang2015phonon}. Data are expressed in\,cm$^{-1}$.}
		\label{tab:exp}
	\end{table}
	Since $E''$ and $A''_2$ are observed both on Ag(111) and Ag(110), we deduce that the interaction of MoS$_2$ with both surfaces is strong enough to activate these modes. 
	The $E'$ and $A'_1$ modes of MoS$_2$/Ag(111) are observed at 377 and 400\,cm$^{-1}$, respectively. In comparison with freestanding-like ML-MoS$_2$ (see blue lines in Figure~\ref{fig:raman}B), the two modes downshift by 6 and 3\,cm$^{-1}$, respectively. 
	The intensity of $A'_1$ is much lower than that of $E'$, with an intensity ratio of $A'_1$ over $E'$ of 0.4.
	This is another anomaly in comparison to the typical Raman spectrum of freestanding-like ML-MoS$_2$ in which the intensity of $A'_1$ is typically larger than that of $E'$~\cite{carvalho2015symmetry, carvalho2016erratum}.
	
	The Raman spectrum of MoS$_2$/Ag(110) is shown in the bottom panel of Figure~\ref{fig:raman}B.
	The most noticeable difference in comparison to MoS$_2$/Ag(111) is the structure of the $A'_1$ mode. 
	The $A'_1$ spectral region of MoS$_2$/Ag(110) shows a strongly asymmetric feature which is accurately fit by two contributions, suggesting that this mode is split into two peaks.
	The one at lower frequency (labeled as ${A'_1}^-$ in Figure~\ref{fig:raman}) is centered at 391.6\,cm$^{-1}$, while the other (${A'_1}^+$) is found at $\sim$400\,cm$^{-1}$.
	The other modes, i.e. $E''$, $E'$ and $A''_2$, are centered approximately at the same positions as in MoS$_2$/Ag(111), as also shown by the scatter plot in Figure~\ref{fig:raman}C.
	However, their Raman lines are narrower, with FWHM being roughly half the FWHM in MoS$_2$/Ag(111) (Figure~\ref{fig:raman}B and Table~\ref{tab:exp}).
	It is known that the linewidth of Raman modes increases for decreasing domain size~\cite{mignuzzi2015effect}.  
	Therefore the narrower linewidths in MoS$_2$/Ag(110) may be related to the larger average size of MoS$_2$ islands on Ag(110) observed with STM (Figure~\ref{fig:stm}B).

	\subsection{Theoretical Modeling of MoS$_2$/Ag Raman Response}
\begin{figure}[th]
		\centering
		\includegraphics[width=.9\textwidth]{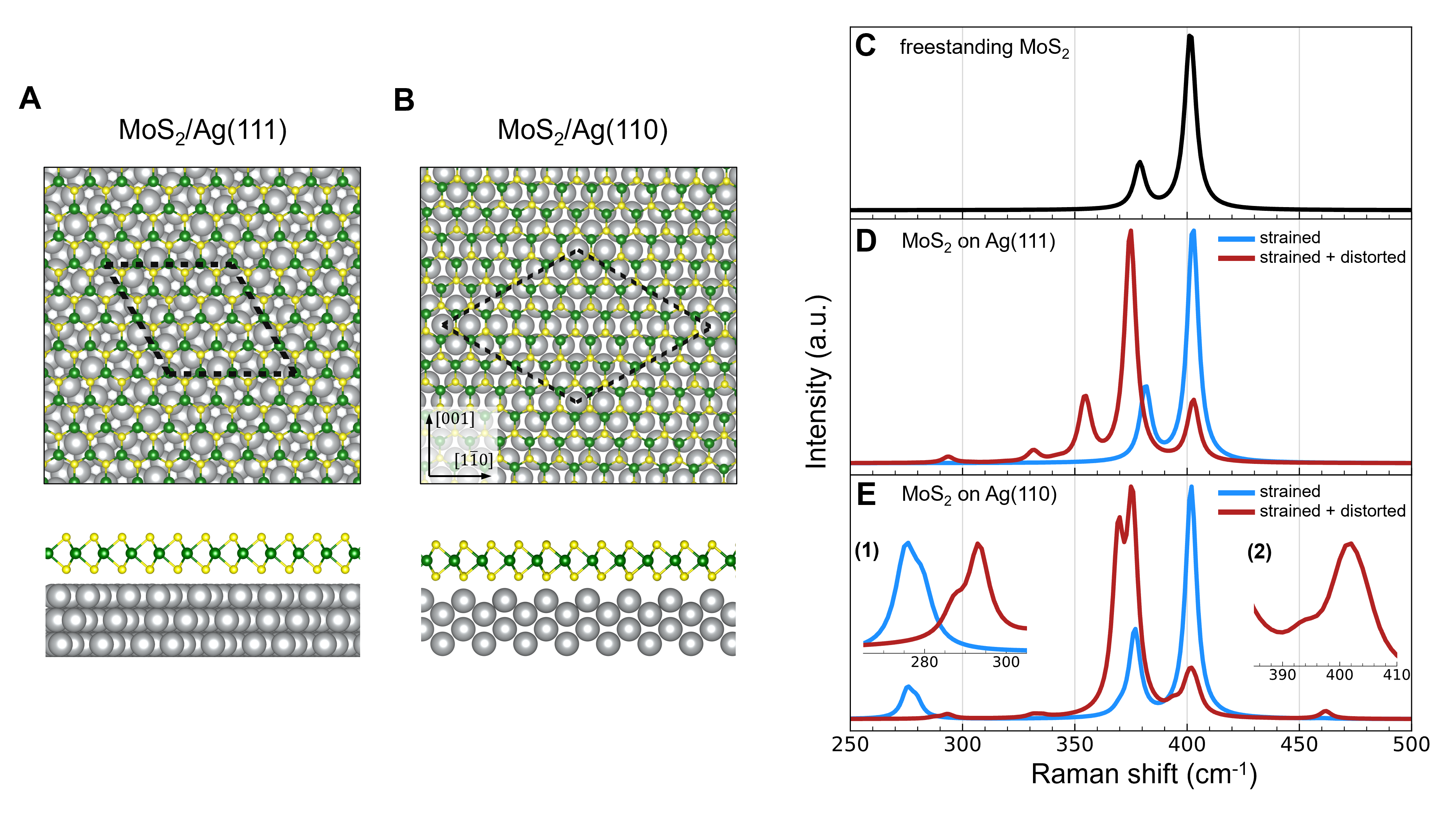}
		\caption{Theoretical modeling of monolayer MoS$_2$ on Ag(111) and Ag(110). (A-B) Top and side views of the ball-and stick models of (A) MoS$_2$/Ag(111) and (B) MoS$_2$/Ag(110). Ag atoms are colored in grey, Mo atoms in green and S atoms in yellow. The dashed black lines outline the coincidence supercells, while the arrows indicate the crystallographic directions of the substrate. (C-E) Calculated Raman spectra of (C) freestanding MoS$_2$, (D) MoS$_2$ on Ag(111) and (E) MoS$_2$ on Ag(110). All the curves are normalized to their maximum intensity. The blue spectra in D and E have been obtained from the strained model (model (i) in the text), which applies only the resulting in-plane strain to a freestanding MoS$_2$ layer. The red spectra have been obtained from the strained-and-distorted model (model (ii) in the text), in which a freestanding MoS$_2$ layer is subjected to both the in-plane strain and the out-of-plane distortion caused by the interaction with the metal surface. Insets E(1) and E(2) show magnified views of the $E''$ and $A'_1$ spectral regions, respectively. The two curves in the E(1) inset have been normalized to their maximum intensity.}
		\label{fig:theory}
	\end{figure}

	We complemented our experimental work with a theoretical investigation of MoS$_2$/Ag interfaces, focused on the calculation of the frequency and intensity of MoS$_2$ Raman modes. 
	The MoS$_2$/Ag(111) interface was modeled with a pseudomorphic cell, where a $4\times4$ supercell of MoS$_2$ is superimposed on a $\sqrt{19}\times\sqrt{19}\;\text{R}\,23.4\degree$ cell of Ag(111). 
	The relaxed structure is reported in Figure~\ref{fig:theory}A. A residual compressive strain, as small as $-0.4$\%, was applied to the MoS$_2$ lattice to adapt to the substrate. 
	The ionic positions are then relaxed as discussed in the Methods section.
	The advantage of using a pseudomorphic cell is to drastically reduce the computational burden required to model the interface as the coincidence lattice observed in the experiment, i.e. $10\times10$ MoS$_2$ on $11\times11$ Ag. This enables the calculation of the vibrational frequencies while still ensuring a reasonably small residual strain.
	A similar approach has been used in the past to study the MoS$_2$/Au(111) interface~\cite{bruix2016single}, with results comparable to those obtained with a realistic model of the moiré coincidence~\cite{tumino2020nature}. 
	Table~\ref{tab:theor_interface} reports the calculated structural and electronic properties associated to the interface between MoS$_2$ and the Ag surfaces.
	For the pseudomorphic MoS$_2$/Ag(111) interface (Table~\ref{tab:theor_interface}, first row), we obtained an adhesion energy of $-60.1\,\text{meV/\AA}^2$. The interfacial dipole moment is negligible ($-0.06\,|\text{e}|\times\text{\AA}$), and so is the change in the metal work function (0.02\,eV). A small charge transfer ($-68.9$ millielectrons per formula unit) takes place from the Ag substrate to the MoS$_2$ layer. A mean average interfacial distance of 258\,pm separates MoS$_2$ from the Ag(111) surface. The close-packed Ag surface maintains an almost flat morphology, with a negligible mean rumpling of 0.89\,pm. The mean rumpling on the MoS$_2$ layer is even smaller, with 0.46\,pm and 0.2\,pm for the bottom and top S layers, respectively.
	
	\begin{table}[ht]
		\centering
		\begin{tabular}{| l | c | c | c | c | c | c | c | c |} 
			\hline
			\textbf{Support} & Strain & $E_{\text{ads}}$ & Dip. & $\Delta\Phi$ & Q/MoS$_2$ & D & R$_{\text{Ag}}$ & R$_{\text{MoS}_2}$\\
			\hline 
			Ag(111) & $-0.4$\% B & $-60.1$ & $-0.06$ & 0.02 & $-68.9$ & 258 & 0.89 & 0.46, 0.2\\ 
			\hline
			Ag(110) & 1.5\% U & $-61.5$ & 0.62 & 0.07 & $-67.1$ & 236 & 2.92 & 1.91, 1.84\\
			\hline
		\end{tabular}
		\caption{Calculated strain (B = biaxial, U = uniaxial), adhesion energy ($E_{\text{ads}}$, meV/\AA$^2$), dipole moment (Dip., $e \times \text{\AA}$), work function change ($\Delta\Phi$, eV), charge transferred to the MoS$_2$ film (Q/MoS$_2$, $10^{-3}|e|$/f.u.), mean interfacial distance (D, pm), mean rumpling on the topmost metallic layer (R$_{\text{Ag}}$, pm), mean rumpling on the MoS$_2$ layer (R$_{\text{MoS}_2}$, pm). The two values of the MoS$_2$ rumpling correspond to the mean rumpling of the bottom and top S layers, respectively. The data have been obtained from the interface models of monolayer MoS$_2$ on Ag(111) and Ag(110) illustrated in Figure~\ref{fig:theory}A-B.}
		\label{tab:theor_interface}
	\end{table}
	The interface of MoS$_2$ with the Ag(110) surface was modeled using a MoS$_2$$-(3\times8)\times\sqrt{2}/(4\times9)-$Ag(110) coincidence (Figure~\ref{fig:theory}B), following the model proposed by Bignardi et al.~\cite{bignardi2021anisotropic} mentioned earlier. A residual uniaxial strain of $+1.5$\% (tensile) is applied along the $[1\bar10]$ direction, while it is negligible along the two other high-symmetry directions of MoS$_2$ to adapt to the metal substrate (Table~\ref{tab:theor_interface}, second row). The adhesion energy is slightly larger compared to Ag(111), $-61.5\,\text{meV/\AA}^2$. A rather large positive dipole moment of $+0.62\,|\text{e}|\times\text{\AA}$ is established at the interface. The change in work function remains quite small ($+0.07$\,eV), as well as the charge transferred to the dichalcogenide film ($-67.1$ millielectrons per formula unit). The interfacial distance (236\,pm) is, however, remarkably smaller compared to the case of Ag(111), and the silver topmost layer undergoes a larger mean rumpling of 2.92\,pm. The rumpling of the MoS$_2$ layer is also significantly larger, with 1.91\,pm and 1.84\,pm for the bottom and top S layers, respectively.   
	
	\begin{table}[ht]
		\centering
		\begin{tabular}{| m{3.5cm} | l | c | c | c | c |} 
			\hline
			\textbf{System} & \textbf{Model} & $E''$ & $E'$ & $A'_1$ & $A''_2$ \\
			\hline 
			Freestanding MoS$_2$ & Relaxed & 280 & 379 & 401 & $\quad$465$\quad$ \\
			\hline
			\multirow{3}{4cm}{MoS$_2$ on Ag(111)} & Strained & 281 & 381 & 403 & 467\\
			& Strained and distorted & 280 & 375 & 403 & 466 \\ 
			& MoS$_2$/Ag slab & 276 & 377 & 397 & 451 \\
			\hline
			\multirow{3}{4cm}{MoS$_2$ on Ag(110)} & Strained & 275*, 280 & 370*, 377 & 402 & 460\\
			& Strained and distorted & 287*, 293 & 369*, 376 & 393, 404 & 462 \\ 
			& MoS$_2$/Ag slab & 284*, 292 & 362*, 367 & 397, 407 & 447 \\
			\hline
		\end{tabular}
		\caption{Calculated vibrational frequencies of the $E''$, $E'$, $A'_1$ and $A''_2$ modes of monolayer MoS$_2$. For MoS$_2$ on Ag surfaces we used three different models that are described in the text. Data are expressed in\,cm$^{-1}$. *These modes are aligned along the stretched lattice direction, i.e. parallel to the $[1\bar10]$ direction of Ag(110).}
		\label{tab:theor_frequencies}
	\end{table}
	Table~\ref{tab:theor_frequencies} reports the calculated vibrational frequencies for the relaxed  freestanding ML-MoS$_2$ (first row) and for three types of models that we used to calculate the vibrational frequencies of Ag-supported MoS$_2$: (i) a strained model, which consists of a freestanding MoS$_2$ monolayer whose lattice is strained as in the interface with Ag, but the ionic positions were relaxed before calculating the frequencies; (ii) a strained-and-distorted model, which consists of a freestanding MoS$_2$ monolayer whose lattice parameters and ionic positions match those of the corresponding metal-supported film; and (iii) the actual MoS$_2$/Ag slabs shown in Figure~\ref{fig:theory}A-B.
	This approach permits to decouple the effect of the lattice strain on the main vibrational modes (strained model), the effects of the out-of-plane structural distortion induced by the interaction with silver (strained and distorted model) and the effect of the charge transfer at the interface (actual ML-MoS$_2$/Ag interfaces). 
	The effect on the vibrational frequencies of biaxial strain, i.e. a deformation expanding or compressing the MoS$_2$ lattice without perturbing the symmetry of the structure, is shown in Figure S3. As expected, a compressive strain, which reduces the Mo-S distances, induces a blue shift in the vibrational frequencies, while a red shift is observed when a tensile strain is applied. Interestingly, the in-plane $E'$ and $E''$ modes are more affected than the out-of-plane $A'_1$ and $A''_2$ modes. As a consequence, the $(A'_1 - E')$ split is strain-dependent.
	Therefore the well-known relation between the frequency difference and the number of MoS$_2$ layers can not be used to infer the Mo$_2$ thickness in the presence of strain.
	
	Figure~\ref{fig:theory}C-E show the simulated Raman spectra for the relaxed freestanding MoS$_2$ (C), MoS$_2$ on Ag(111) (D), and MoS$_2$ on Ag(110) (E). The calculation of the intensity for the MoS$_2$/Ag slabs (i.e. model (iii) introduced before) presents severe issues of feasibility, related to the difficulty of calculating the electronic polarizability on a conductive system, and to the extremely high computational burden. Therefore, we report in Figure~\ref{fig:theory}D-E only the spectra calculated using the strained (blue lines) and the strained-and-distorted (red lines) models.
	This fact clearly limits the possibility of a 1:1 comparison with the experimental spectra. 
	The freestanding MoS$_2$ monolayer presents four vibrational modes falling at 280\,cm$^{-1}$ ($E''$), 379\,cm$^{-1}$ ($E'$), 401\,cm$^{-1}$ ($A'_1$) and 465\,cm$^{-1}$ ($A''_2$), Table~\ref{tab:theor_frequencies}. Only two modes display a significant intensity, namely $E'$ and $A'_1$ (most intense) (Figure~\ref{fig:theory}E), whereas $E''$ and $A''_2$ and not Raman active.
	
	On Ag(111), where the pseudomorphic interface model (Figure~\ref{fig:theory}A) introduces a small biaxial compressive strain in MoS$_2$ (Table~\ref{tab:theor_interface}), a blue shift of 2\,cm$^{-1}$ was observed on all normal modes in the strained model (Table~\ref{tab:theor_frequencies}), with almost no effect on their relative intensity (Figure~\ref{fig:theory}D, blue line). From a comparison with the strained-and-distorted model, we can argue that the main effect related to the out-of-plane distortion at the interface with Ag(111) is the red shift of the $E'$ mode at 375\,cm$^{-1}$ (Table~\ref{tab:theor_frequencies}). Notably, the $A'_1$ and $E'$ modes undergo an inversion in their relative intensities (Figure~\ref{fig:theory}D, red line). 
	This effect is in agreement with the experimentally observed inversion of the intensity ratio that characterizes the Raman spectrum of metal-supported MoS$_2$ (Figure~\ref{fig:raman}B), suggesting that the out-of-plane distortion induced by the interaction with the metal significantly contributes to suppressing the $A'_1$ intensity relative to $E'$.
	Satellite peaks corresponding to mixed modes with non-negligible intensities are also present in the simulated spectrum (Figure~\ref{fig:theory}D, red line) in the 280--360\,cm$^{-1}$ region. These peaks do not appear in the experimental spectrum (Figure~\ref{fig:raman}B, top panel), suggesting that they may be related to artifacts introduced by the absence of the metal support in the strain-and-distorted model.
	The $A''_2$ mode remains silent in the simulated spectra, contrarily to what observed experimentally. 
	Note, however, that previous experimental works~\cite{scheuschner2015interlayer, lee2015anomalous} have found that the $A''_2$ intensity can be enhanced by excitonic resonance effects which are not considered by our theoretical calculations. 
	In the MoS$_2$/Ag(111) slab model, the presence of the charge transfer and interfacial dipole induces a red shift of $E''$ to 276\,cm$^{-1}$, and a small blue shift of $E'$ to 377\,cm$^{-1}$ (Table~\ref{tab:theor_frequencies}). The $A'_1$ mode is also slightly red-shifted at 397\,cm$^{-1}$, while $A''_2$ is more noticeably red-shifted to 451\,cm$^{-1}$. 
	These frequencies are in good agreement with our experimental data (Table~\ref{tab:exp}).  
	
	On Ag(110), the mostly uniaxial strain undergone by the MoS$_2$ layer induces some important changes in the vibrational features. In the strained model, the $E''$ mode splits at 275 and 280\,cm$^{-1}$ (Table~\ref{tab:theor_frequencies}) and becomes Raman active (Figure~\ref{fig:theory}E and E(1) inset, blue line). A similar splitting at 370\,cm$^{-1}$ and 377\,cm$^{-1}$ is observed for $E'$, albeit less distinguishable in the calculated spectrum. Notably, for both $E'$ and $E''$, the low frequency component corresponds to atomic motions aligned to the strained lattice direction, i.e. parallel to the $[1\bar10]$ direction of the substrate. While $A'_1$ does not show any relevant frequency shift, the $A''_2$ mode is red-shifted at 460\,cm$^{-1}$ with respect to the freestanding monolayer. 
	The strained-and-distorted model predicts that the structural distortion induced by the interaction with the Ag(110) substrate has also important effects. 
	The split $E''$ mode is blue-shifted at 287 and 293\,cm$^{-1}$, whereas $E'$ is slightly red-shifted with respect to the strained model (Table~\ref{tab:theor_frequencies}). The $E'$ split is clearly visible in the more intense spectrum obtained from the strained-and-distorted model (Figure~\ref{fig:theory}E, red line).
	In this model, the $A'_1$ mode is also split at 393 and 404\,cm$^{-1}$, with a frequency gap of 11\,cm$^{-1}$ (Figure~\ref{fig:theory}E, inset E(2)). The $A''_2$ mode falls at 462\,cm$^{-1}$, slightly blue-shifted with respect to the strained case. 
	Notably, both $E''$ and $A''_2$ are Raman active, even though with a rather small intensity. 
	In the MoS$_2$/Ag(110) slab model, the presence of the metal surface slightly modify the position and split of the $E''$, $E'$ and $A'_1$ modes (Table~\ref{tab:theor_frequencies}). The $A''_2$ frequency decreases to 447\,cm$^{-1}$, in better agreement with the experimental value of 452.9\,cm$^{-1}$.
	
	The split of the $E'$ mode predicted by our calculations agrees with previous experimental works~\cite{rice2013raman, conley2013bandgap} on strained ML-MoS$_2$, which have shown that a uniaxial strain lifts the degeneracy of $E'$ and splits this mode into two subpeaks. 
	Nevertheless, we do not detect any appreciable splitting of the $E'$ mode in the experimental Raman spectrum of ML-MoS$_2$/Ag(110) (Figure~\ref{fig:raman}B, bottom panel).
	It is possible, however, that the calculations overestimate the frequency split and that the real value is too small to produce a detectable splitting of the experimental peak.
	On the other hand, the splitting of the $A'_1$ mode is observed in both experiments and calculations. Although the calculated frequency of the two submodes (i.e. 397 and 407\,cm$^{-1}$ for the MoS$_2$/Ag(110) slab model in Table~\ref{tab:theor_frequencies}) is blue-shifted compared to the experimental values (391.6 and 400.5\,cm$^{-1}$, Table~\ref{tab:exp}), the frequency split is in very good agreement.
	Notably, the calculated splitting is not predicted from the strained model but only from the strained-and-distorted and the MoS$_2$/Ag(110) slab models (Table~\ref{tab:theor_frequencies}).
	This result evidences that the MoS$_2$ Raman response is strongly influenced by the out-of-plane distortion resulting from the interaction with the metal surface.

	\subsection{Decoupling MoS$_2$ from the Silver Surface}
\begin{figure}[htb!]
		\centering
		\includegraphics[width=.75\textwidth]{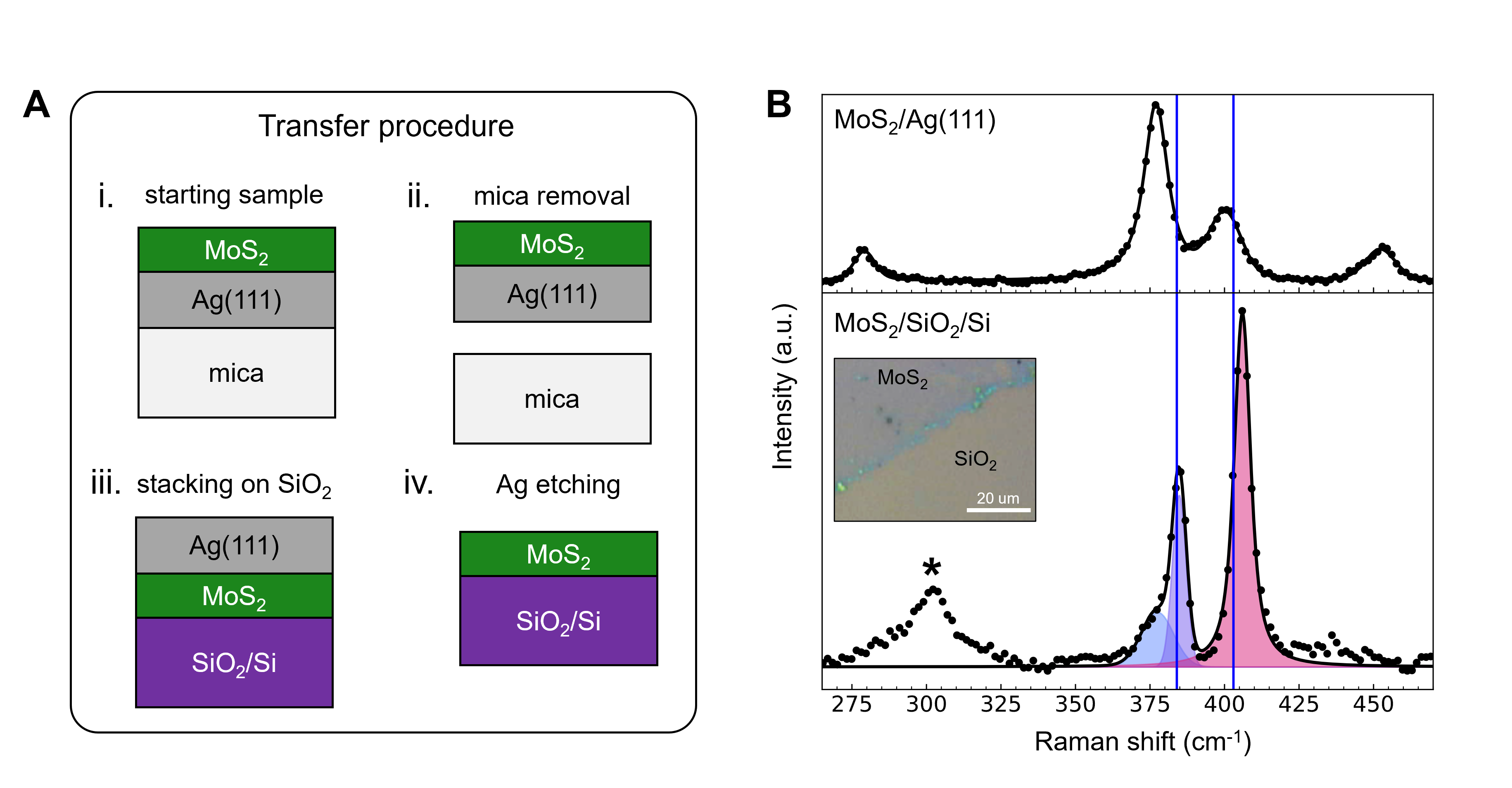}
		\caption{Transfer of MoS$_2$ from Ag(111)/mica onto a silica substrate. (A) Schematics of the transfer procedure. From the initial sample (i), we cleaved the mica substrate off and placed MoS$_2$/Ag on top of the SiO$_2$/Si substrate, with MoS$_2$ sandwiched between SiO$_2$ and Ag (iii). Finally, the Ag film was etched away using a potassium iodide solution (iv). (B) Raman spectra of ML-MoS$_2$ on Ag(111) (top) and after transfer onto SiO$_2$/Si (bottom). Experimental points are indicated by black dots, peak-fitting functions by colored areas, the total fit by a solid black line. The asterisk marks a feature related to the Si substrate. The two vertical blue lines mark the $E'$ and $A'_1$ peak positions of MoS$_2$/Ag(111). Inset: Optical microscopy image showing the contrast between the transferred MoS$_2$ layer and the bare SiO$_2$ surface.}
		\label{fig:transfer}
	\end{figure}

	We discussed both experimentally and theoretically how the vibrational properties of ML-MoS$_2$ are strongly influenced by the interaction with the supporting metal surface.
	To provide further evidence we decoupled MoS$_2$ from the Ag surface by transferring it onto a SiO$_2$/Si substrate. 
	The transfer procedure is depicted in Figure~\ref{fig:transfer}A and described in the Methods section. 
	Briefly, we first peeled MoS$_2$/Ag(111) off of mica (Figure~\ref{fig:transfer}A.i), then we stacked it onto the new substrate with MoS$_2$ facing the SiO$_2$ surface (Figure~\ref{fig:transfer}A.ii), and finally we etched the Ag film using a potassium iodide solution (Figure~\ref{fig:transfer}A.iv). 
	The resulting sample was inspected via optical microscopy. The 300\,nm thickness of the SiO$_2$ film maximizes the optical contrast of MoS$_2$, making it easily distinguishable from the bare SiO$_2$ surface (inset of Figure~\ref{fig:transfer}B, bottom panel). 
	The Raman spectrum of MoS$_2$/SiO$_2$ is shown in the bottom panel of Figure~\ref{fig:transfer}B, in comparison with the spectrum of MoS$_2$/Ag(111) (top panel). The two vertical lines mark the typical $E'$ and $A'_1$ peak positions of freestanding-like MoS$_2$ reported in previous works~\cite{conley2013bandgap, li2012bulk, zhang2015phonon}. 
	Overall the spectrum taken after transfer looks much more similar to the typical spectrum of ML-MoS$_2$ on a weakly interacting substrate.
	By comparing it with the original spectrum of MoS$_2$/Ag(111) we notice three major differences. 
	First, the $E'$ and $A'_1$ modes shift up to 384.8 and 406\,cm$^{-1}$, respectively. 
	There is a minor component at 377\,cm$^{-1}$, aligned with the $E'$ mode in MoS$_2$/Ag(111), which can be due to residual Ag inducing local strain on MoS$_2$, or to disorder in the MoS$_2$ lattice~\cite{mignuzzi2015effect}. A disorder induced broadening of the $A'_1 - E'$ frequency gap, previously reported~\cite{mignuzzi2015effect}, may also explain why the measured 21.2\,cm$^{-1}$ difference is slightly above the commonly reported 18-20\,cm$^{-1}$ values for freestanding-like monolayer MoS$_2$.  
	Second, the intensity of $A'_1$ increases dramatically relative to $E'$, leading the intensity ratio $A'_1$/$E'$ to increase from 0.4 to 2.
	Third, the $E''$ and $A''_2$ modes disappear (the feature marked with an asterisk is related to the silicon substrate). 
	These three observations provide further evidence for the strong sensitivity of MoS$_2$ Raman response to the interaction with metals. Releasing such interaction by decoupling the MoS$_2$ layer from the metal surface restores a more freestanding-like behavior of the vibrational modes.

	
	\section{Conclusions}
	We studied the Raman response of monolayer MoS$_2$ grown on Ag(111) and Ag(110) under UHV conditions. 
	The experimental Raman spectra show significant deviations from the freestanding MoS$_2$ behavior, both in the position and intensity of the observed active modes. 
	By comparing the spectra of MoS$_2$/Ag(111) and MoS$_2$/Ag(110), we observed a notable difference in the out-of-plane $A'_1$ vibrational mode, which on Ag(110) splits into two peaks $\sim$9\,cm$^{-1}$ apart. 
	This splitting is corroborated by our theoretical calculations, from a model which includes both the in-plane strain and the out-of-plane distortion of the MoS$_2$ lattice induced by the metal support.    
	Our results provide further insight into the sensitivity of MoS$_2$ Raman response to the metal surface structure, and support the characterization and implementation of MoS$_2$ in electronic devices.
	
	
	
	\begin{acknowledgement}
		A.L.B. and P.D.A. acknowledge PRIN 2022 project 2022XMYF5E ``Intercalation assisted silicene/MoS$_2$ heterostructures for two-dimensional nanojunctions'' funded by the Italian Ministry for University and Research (MUR).
		
		V.R., A.L.B., and C.S.C acknowledge funding by: 
		- Funder: project funded under the National Recovery and Resilience Plan (NRRP), Mission 4 Component 2 Investment 1.3 - Call for tender No. 341 of 15.03.2022 of Ministero dell'Università e della Ricerca (MUR); funded by the European Union NextGenerationEU    
		- Award Number: project code PE0000021, Concession Decree No. 1561 of 11.10.2022 adopted by Ministero dell'Università e della Ricerca (MUR), CUP D43C22003090001, Project title ``Network 4 Energy Sustainable Transition\textunderscore NEST''.
				
		S.T. acknowledges funding by: - ICSC – Centro Nazionale di Ricerca in High Performance Computing, Big Data and Quantum Computing, funded by European Union - NextGenerationEU
		- PRIN Project 2022LS74H2 funded by the Italian Ministry for Universities and Research (MUR)
		- PRIN Project P20227XSAH by the Italian Ministry for Universities and Research (MUR), in the context of the National Recovery and Resilience Plan and co-financed by the Next Generation EU.
	\end{acknowledgement}
	
	\begin{suppinfo}
		
		Supplemental information are available: additional description of MoS$_2$/Ag interface geometries, supplemental STM and calculated Raman data.
		
	\end{suppinfo}
	
	
	\bibliography{mos2_references}
	
\end{document}